\documentclass[iop]{emulateapj}

\usepackage{graphicx}

\def\etal{{\it et al.}}
\def\msun{M$_\odot$}

\shorttitle{Mock Blue Straggler Observations}
\shortauthors{Sills \etal}

\begin{document}
\title{Mock Observations of Blue Stragglers in Globular Cluster Models}
\author{Alison Sills}
\affil{Department of Physics and Astronomy, McMaster University, Hamilton, Ontario L8S 4M1, Canada}
\email{asills@mcmaster.ca}
\author{Evert Glebbeek\altaffilmark{1}}
\affil{Department of Astrophysics/IMAPP, Radboud University Nijmegen, PO Box 9010, 6500 GL, Nijmegen, The Netherlands} 
\email{e.glebbeek@astro.ru.nl}
\author{Sourav Chatterjee}
\affil{Department of Astronomy, University of Florida, Gainesville, FL 32611, USA}
\email{s.chatterjee@astro.ufl.edu}
\author{Frederic A. Rasio}
\affil{Center for Interdisciplinary Exploration and Research in Astrophysics (CIERA) and Department of Physics and Astronomy, Northwestern University, 2145 Sheridan Road, Evanston, IL 60208, USA}
\email{rasio@northwestern.edu}

\altaffiltext{1}{Department of Physics and Astronomy, McMaster University, Hamilton, Ontario L8S 4M1, Canada}

\begin{abstract}

We created artificial color-magnitude diagrams of Monte Carlo dynamical models of globular clusters, and then used observational methods to determine the number of blue stragglers in those clusters. We compared these blue stragglers to various cluster properties, mimicking work that has been done for blue stragglers in Milky Way globular clusters to determine the dominant formation mechanism(s) of this unusual stellar population. We find that a mass-based prescription for selecting blue stragglers will choose approximately twice as many blue stragglers than a selection criterion that was developed for observations of real clusters. However, the two numbers of blue stragglers are well-correlated, so either selection criterion can be used to characterize the blue straggler population of a cluster. We confirm previous results that the simplified prescription for the evolution of a collision or merger product in the BSE code overestimates their lifetimes. We show that our model blue stragglers follow similar trends with cluster properties (core mass, binary fraction, total mass, collision rate) as the true Milky Way blue stragglers, as long as we restrict ourselves to model clusters with an initial binary fraction higher than 5\%.  We also show that, in contrast to earlier work, the number of blue stragglers in the cluster core does have a weak dependence on the collisional parameter $\Gamma$ in both our models and in Milky Way globular clusters. 
\end{abstract}


\section{Introduction}

Blue stragglers are main sequence stars that are brighter and bluer than the main sequence turnoff in their environment. They are found in open and globular clusters \citep{deMarchi06, Moretti08}, in dwarf galaxies \citep{Mapelli07}, and even in the field \citep{Preston00}. Since most of these environments do not contain enough gas to support a current or recent burst of star formation, the expectation is that blue stragglers are formed through some interaction which adds mass to a normal main sequence stars. The two dominant formation mechanisms are expected to be stellar collisions and binary mass transfer. In dynamically active environments such as globular clusters, we also expect that both of these mechanisms could be moderated by dynamical interactions. For example, collisions can occur during close interactions between pairs of binary stars; close encounters could also modify binary orbits so that mass transfer happens either sooner or later than would occur in an unperturbed situation. Despite many studies to try to disentangle these effects, we have not been able to convincingly determine which process(s) are dominant in globular clusters.

Based on calculations of predicted collision rates, it was expected that collisions were most important in the cores of GCs \citep[e.g.][]{Leonard89}. There was even evidence from dynamical models using a static cluster background that collisions dominated in the cores while binary coalescence was more important in the outskirts \citep{Mapelli06}. However, a survey of blue stragglers in globular clusters found no correlation between the fraction of blue stragglers and the collision rate in clusters \citep{Davies04}, and in fact the two quantities are slightly anti-correlated. A more detailed look at the same survey \citep{Knigge09} confirmed the lack of correlation between blue straggler number and collisional properties, and found that the best correlation was in fact with core mass. The inference is that the binary population of the cluster (which scales with core mass) is the dominant driver of blue straggler population in a cluster. 

At the same time, various groups were modeling populations of blue stragglers in clusters. The first large-scale N-body model to include stellar and binary evolution and a population of primordial binaries, and to specifically look at the resultant blue straggler population, was the work of \citet{Hurley05} for the open cluster M67. He found that both dynamics and binary evolution were important in creating the present-day blue straggler population. Similar models were calculated for a slightly older open cluster, NGC 188 \citep{Geller13}. Particular attention was paid to the initial conditions for the binary population, and the authors find that the models underproduced blue stragglers formed via mass transfer. They suggest that the criteria for invoking a common envelope phase during mass transfer may be too strict. Since direct star-by-star N-body models for globular clusters are still prohibitively expensive, \citet{Leigh11b} used a simple analytic prescription for binary and single star encounter rates and the expected number of binary stars. They found that in order to match the observed numbers of blue stragglers in globular clusters, binary evolution should dominate over collisional processes.  Most recently, Monte Carlo dynamical models of globular cluster evolution have started to include stellar and binary evolution, and therefore can follow the creation of blue stragglers \citep{Hypki13, Chatterjee13}

To date, however, there has been a fundamental disconnect between the theoretical and observational investigations of blue stragglers in clusters. Theoretical blue stragglers are identified by their mass compared to the turnoff mass. Observational blue stragglers are chosen simply by their position in the color-magnitude diagram of a cluster. Many of the studies quoted above chose the blue stragglers in only one color band. That band is typically V-I, which is not ideal for selecting hot stellar populations as shown by \citet{Ferraro97}. Therefore, there is always an uncertainty in the observed blue straggler populations -- are they all really main sequence stars, more massive than the turnoff, or are we contaminated by chance superpositions, blends of binary stars, and photometric errors or anomalous populations? \citet{Sills00} showed that even using three photometric bands (U, B and V) to select blue stragglers in 47 Tucanae resulted in 8\% of the objects from the B-V color-magnitude diagram to be rejected as true blue stragglers, since they were not in the correct part of the color-magnitude diagram in both U-B and B-V colors. 

We can now address this issue, however, by looking at the populations of model blue stragglers in the suite of cluster models from Monte Carlo modeling. In this paper, we ``observe" the blue stragglers in the simulated clusters presented in \citet{Chatterjee13} (Paper I). We determine the blue stragglers and other cluster properties from the color-magnitude diagrams, using methods that are as close as possible to those used by the various observational groups who do this for real clusters. We wish to determine if there are any observational biases which affect the conclusions from those groups. We also wish to determine if the theoretical models have any short-comings that renders any comparison to observations invalid. 

In section 2, we outline our method for ``observing" the blue stragglers in the model clusters. Section 3 gives detailed comparisons between our model blue stragglers and a number of significant observational results about blue stragglers from the past decade. In section 4, we present our conclusions and discuss future directions. 

\section{Methods}

A large collection of models of globular clusters was constructed using the Cluster Monte Carlo (CMC) code \citep{Joshi00,Joshi01,Fregeau03,Fregeau07,Chatterjee10,Umbreit12}. These models are described in detail in \citet{Chatterjee13a}. They include may physical processes such as two body relaxation, physical collisions, binary-mediated scattering. Binary and single star evolution is treated using BSE \citep{Hurley02}. A number of initial cluster parameters were explored, including initial binary fraction, initial virial radii, initial number of stars, and initial concentration. The clusters were evolved to an age of 12 Gyr. We have a total of 128 models, which span a range of final properties such as total mass, core mass, and core radius. The blue straggler populations in these models were analyzed in detail in Paper I. In that paper, the selection of blue stragglers was done based on non-observational properties; now we wish to test that method against observational selection.

In order to compare the model blue stragglers with the observed population of blue stragglers in real Milky Way clusters, we converted the stellar luminosities and effective temperatures to HST/ACS F606W and F814W magnitudes. We used the color transformation program of \citet{Dotter08}, modified to include the ACS filters (Dotter, personal communication). These are the same filters that are used in the ACS Survey of Galactic Globular Clusters \citep{Sarajedini07} and therefore are those used by \citet{Leigh11} to select blue stragglers and other stellar populations in a self-consistent way from color-magnitude diagrams. The colors and magnitudes of binary stars are calculated by determining the magnitudes of each component, and then combining them in the appropriate way so that each binary is ``observed" as a single object.

\begin{figure}
\begin{center}
\includegraphics[width=3.2in]{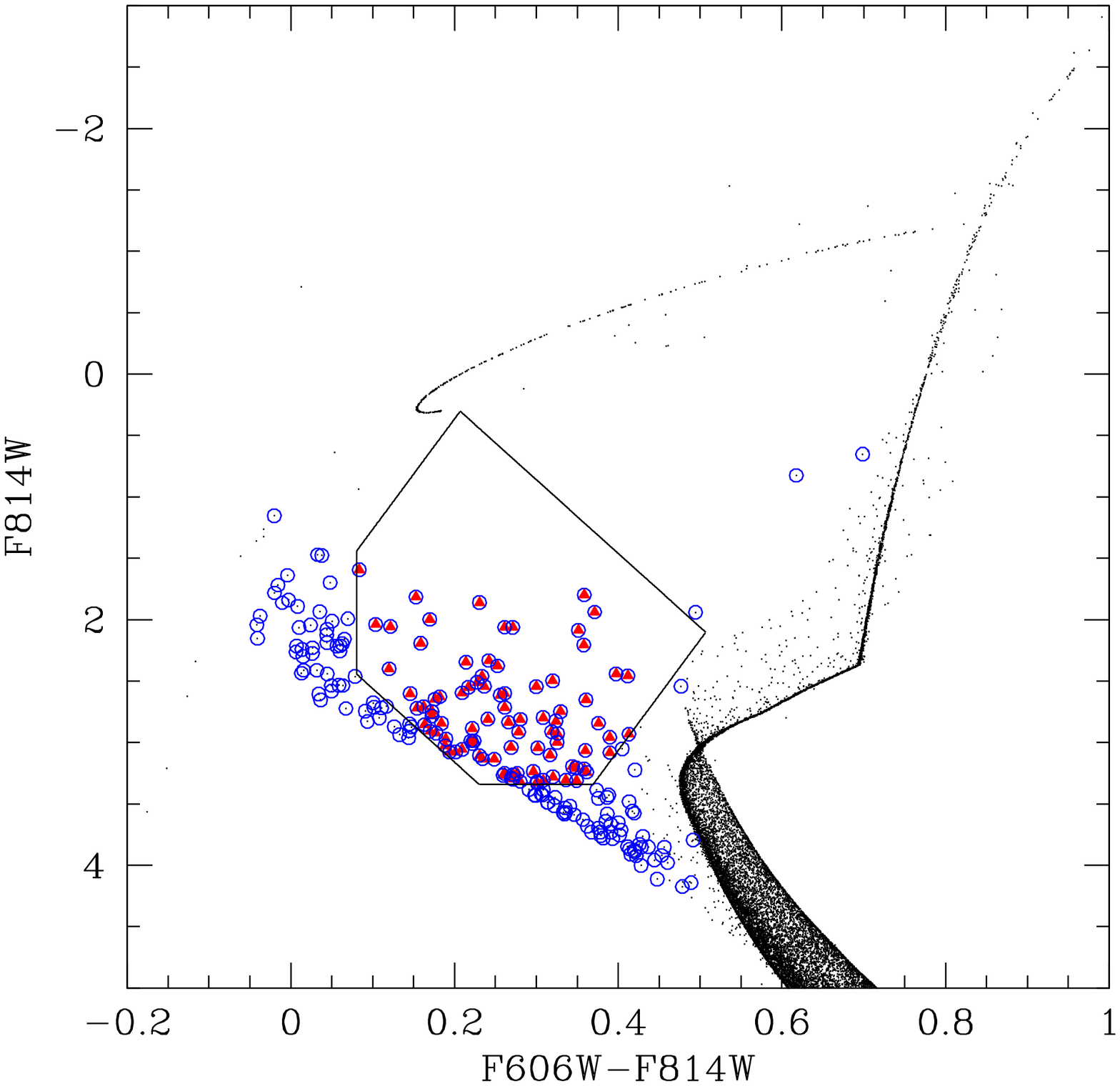}
\end{center}
\caption[]{Color-magnitude diagram of run 110 from \citet{Chatterjee13}. The blue straggler selection box from \citet{Leigh11} is shown, along with the selected blue stragglers as solid triangles, red in the online edition. The objects selected as blue stragglers using the method of Paper I are circled. While the objects within the box are also selected by the theoretical method, the theoretical selection includes more main sequence blue stragglers than would be identified by \citet{Leigh11}. }
\label{fig:CMD}
\end{figure}

We applied the selection criteria of \citet{Leigh11} to our model clusters. These criteria have some free parameters, namely the location of the turnoff and of the horizontal branch, which were determined from the color-magnitude diagrams. A color-magnitude diagram of one of our models is shown in Figure \ref{fig:CMD}. This is the same model cluster whose HR diagram is shown in figure 1 of Paper I. The observational blue straggler selection box is shown, and all blue stragglers selected by this technique are shown as solid triangles. The circled stars are objects which meet the theoretical selection criteria of Paper I: they are main sequence (core-hydrogen burning) stars with masses greater than 1.1 times the current turnoff mass (0.835 \msun in this model cluster). For binary systems, either of the components had to meet these criteria. 

It is encouraging that the majority of the objects inside the selection box are chosen by both methods. In some clusters, we see one or two stars, brighter than the majority of the blue stragglers, which are not selected by the theoretical criteria but are within the selection box. These are stars which do have masses greater than 1.1 times the turnoff mass, but are not core hydrogen burning. They are former blue stragglers who are currently traversing the Hertzsprung gap. Because this phase is short-lived, their contamination of the blue straggler population is small, and one could argue that we should include them in our count because they do trace the blue straggler formation efficiency as much as the true, main sequence blue stragglers. 

The objects outside the observational selection box are also instructive. First, there are blue stragglers fainter than our selection box. These blue stragglers have masses which are only slightly larger than the cutoff mass. These objects are also seen in real color-magnitude diagrams of clusters, but are confused with blends of binary systems (the small dots that are not circled in that region of Figure \ref{fig:CMD} mark the blends of binary systems detected in our models), and any photometric error will also broaden the main sequence turnoff in this region. Removing these objects from any standardized selection of blue stragglers is sensible as the confusion between the three contributors (low mass BSS, stellar blends, and photometric error) is large. 

The model clusters show a group of blue stragglers which are even bluer than the blue straggler selection box. These objects do not appear in the color-magnitude diagrams of the real clusters in the ACS Survey. Their presence in the model color-magnitude diagrams is an artifact of the treatment of mergers within BSE. When two stars merge, BSE assigns a homogeneous composition to the remnant which is based on the total amount of hydrogen and helium present in the two parent stars. This fully mixed product has a long lifetime because hydrogen in brought to the interior, and is blue because of the enhanced surface helium abundance. In some earlier papers in the literature, both collision products \citep{Benz92} and binary coalescence products \citep{Lu10} have been assumed to be fully mixed. More detailed models \citep{Lombardi95,Sills01,Glebbeek08} have shown that in fact collision products retain a strong memory of the chemical profiles of their parents. Models of binary coalescence \citep{Chen08} are also inconsistent with full mixing during the merging process. The lack of real blue stragglers to the blue of the \citet{Leigh11} selection box, combined with their presence in our model clusters, suggests that the detailed models are correct and no formation mechanism produces fully mixed blue stragglers. The error is more severe for more massive blue stragglers, which come from more evolved parents and should therefore have very short main sequence lifetimes \citep{Sills97}. Applying a correction factor as proposed by \citet{Glebbeek08} reduces the lifetime of the merger products. The most massive merger products then evolve away from the main sequence and would no longer appear as blue stragglers in the simulations.

There are also objects that lie above the subgiant branch and between the blue stragglers and the giant branch. These are binary stars, which contain a main sequence star (the blue straggler) plus an RGB star, a subgiant branch star, or a newly-formed helium white dwarf. Their positions in the color-magnitude diagram are dominated by the light of the other object in the system rather than that blue straggler. In the simulation shown in figure \ref{fig:CMD}, the blue straggler - RGB systems are the two objects near the RGB at a magnitude of $\sim$ 1. The subgiant system is just outside the selection box on the bright side, and the object just outside the selection box nearest the turnoff is a blue straggler - helium WD binary. By choosing this specific observational selection box, we are removing legitimate blue stragglers from our consideration because they are in a binary system with something quite bright. Since the numbers of these objects are quite small, we expect that any bias introducted by rejecting these objects as blue stragglers should not be significant.

\section{Results}

In this section, we will compare the number of ``observationally-selected" model blue stragglers (i.e. those in the selection box shown in Figure \ref{fig:CMD}) to a variety of cluster properties, guided by papers that have done this for real clusters over the past decade. The cluster properties (total mass, core mass, etc.) for the model clusters are those calculated as described in \citet{Chatterjee13a}. In particular, the core properties were determined by creating artificial surface brightness profiles and determining the core radius from those, rather than using the standard dynamical definition of core radius. Therefore, we have ``observational" properties for the models which can be directly compared to real clusters. The observed number of blue stragglers are taken from \citet{Leigh11}, and the observed cluster properties are taken from the Harris catalogue \citep[][2010 revision]{Harris96}.

\begin{figure}
\includegraphics[width=3.2in]{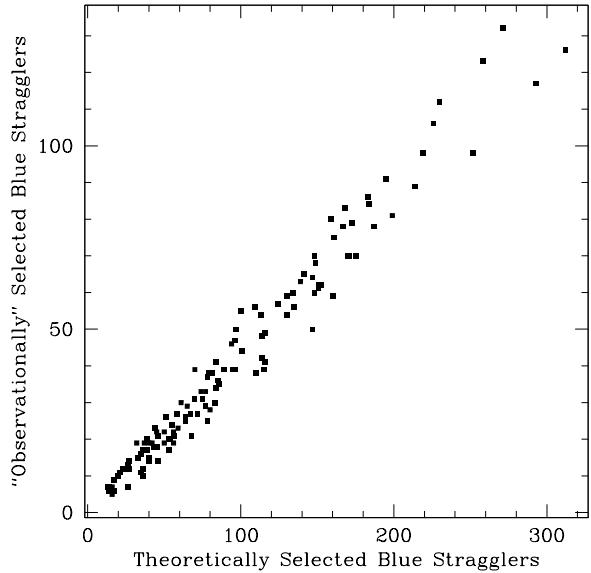}
\caption[]{Number of ``observationally" selected blue stragglers (those inside the selection box in figure \ref{fig:CMD}) compared to theoretically selected blue stragglers (the circled points in figure \ref{fig:CMD}). As expected, the number of theoretically selected blue stragglers is larger, but the two selection methods are very well correlated.}
\label{fig:MethodCompare}
\end{figure}

In Figure \ref{fig:MethodCompare}, we plot the number of model blue stragglers selected in two different ways: theoretically selected numbers on the x-axis vs ``observationally" selected on the y-axis. The theoretical selection method chooses about twice as many blue stragglers as the method which uses the selection box in the color-magnitude diagram, as expected from Figure \ref{fig:CMD}. However, the correlation is quite good. We also confirm that the fraction of observationally-selected blue stragglers to theoretically selected blue stragglers does not depend on any cluster property, such as initial mass, binary fraction, or initial concentration. Therefore, we confirm that the observational selection procedure is robust and can be used to determine population sizes, and certainly can be used to look at correlations between blue straggler populations and cluster properties. 

We looked at the formation history of the blue stragglers inside and outside the observational selection box, to see if the observational box is preferentially choosing blue stragglers made in a particular way. We found that there is no clear bias in the selection procedure. Collisional blue stragglers have, on average, the same fraction in the model clusters using either criterion, with a small scatter. Mass transfer binaries are also selected with approximately the right fraction, although in this case, there is a larger spread from model to model, with a few models having almost all their mass transfer systems inside the observational selection box and others with almost none inside the observational box. We conclude that the observational selection criterion samples the blue stragglers created from all formation channels without any clear bias, if a large enough sample of clusters is considered.

In the following sections, we restrict ourselves to blue stragglers selected using the observational box and found in the core of the clusters, so that we can make a direct comparison to the observed blue stragglers found in \citet{Leigh11}.

\subsection{The blue straggler-cluster property correlations}

\begin{figure}
\begin{center}
\includegraphics[width=3.2in]{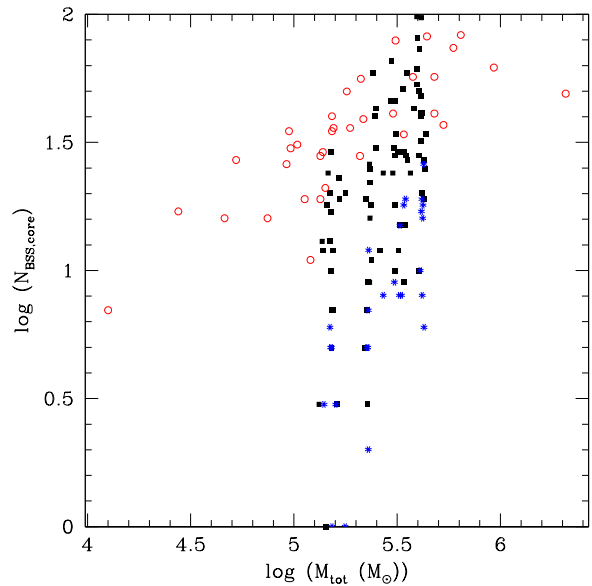}
\end{center}
\caption[]{Number of blue stragglers in the core vs total mass of the
  cluster. The data for model clusters are shown as solid squares, if the initial binary fraction of the model was greater than 5\%, and as stars if the initial binary fraction was equal to 5\%.
  The data for Milky Way clusters are shown as open circles.}
\label{fig:nmtot}
\end{figure}

A leap forward in the study of blue straggler populations came from the HST/WFPC2 survey of globular clusters \citep{Piotto02}. This was the first detailed and self-consistent look at the cores of many globular clusters at once (74 in this case), and a substantial population of blue stragglers was found in every cluster. A subsequent paper \citep{Davies04} tested the prediction that the number of blue stragglers should be correlated with the collision rate in the cluster. As a control, they also compared the number of blue stragglers to the total mass in the cluster. Surprisingly, the correlation with total mass was stronger than the correlation with collision rate, and if anything, there was a weak {\it anti-}correlation with collision rate. These results have subsequently been confirmed by various other authors using the same dataset \citep{Leigh07,Knigge09}.  In Figure \ref{fig:nmtot}, we plot the blue stragglers in the core vs total cluster mass, with the models in solid squares and the observations in open circles. The models have a restricted range of total mass compared to the observations, but in the regions where they overlap, we predict approximately the correct number of core blue stragglers, as long as we restrict ourselves to model clusters where the binary fraction is greater than 5\% (solid squares). In agreement with all previous observations, we have more blue stragglers in more massive clusters. In the next three figures and accompanying analysis, we present only model clusters where the initial binary fraction is 10\% or higher.

\begin{figure}
\begin{center}
\includegraphics[width=3.2in]{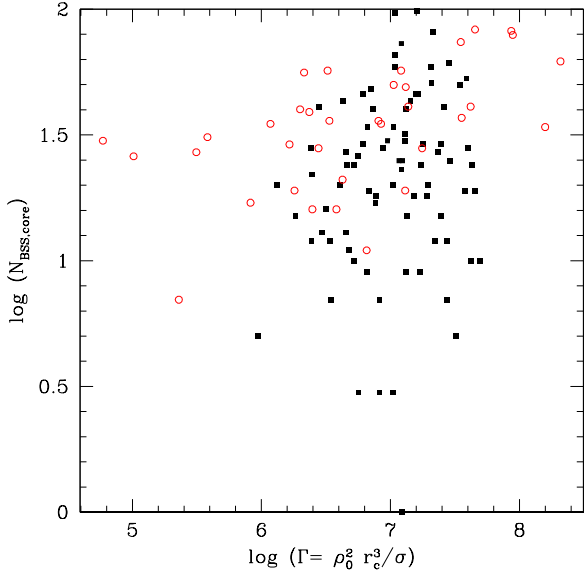}
\end{center}
\caption[]{Number of blue stragglers in the core vs the collision rate
  in the cluster calculated from current cluster properties: central density in $M_{\odot}/pc^3$, core radius in pc, and velocity dispersion in $km/s$. The model
  clusters are shown as solid squares, and the Milky Way clusters are
  shown as open circles.}
\label{fig:ngamma}
\end{figure}

Figure \ref{fig:ngamma} shows the number of core blue stragglers vs the collisional parameter $\Gamma$. Following \citet{Leigh07}, we calculate $\Gamma$ under the assumption that the cluster is well-fit by a King model, so that the collision rate is proportional to $\rho_0^2 r_c^3 / \sigma$, where $\rho_0$ is the central mass volume density, $r_c$ is the core radius, and $\sigma$ is the central velocity dispersion \citep{Pooley06}. Again, while our model clusters span a narrower range in $\Gamma$ than the observations, the model results overlap the observational data. We note that both the models and the Milky Way clusters do show a correlation with $\Gamma$, although not a very strong dependence (slope = 0.15, Spearman correlation coefficient=0.58 for the Milky Way clusters, slope=0.16, Spearman correlation coefficient=0.17 for the models). This is consistent with the correlations shown by \citet{Leigh13} between the number of blue stragglers and the rates of single-single, single-binary, and binary-binary collisions. Earlier papers which claimed an anti-correlation or no correlation between blue stragglers and the global collision rate $\Gamma$ used samples of blue stragglers from the WFPC2 globular cluster survey \citep{Piotto02}. The better photometric data from the ACS survey shows a clearer connection between blue straggler numbers and collisions. A more detailed discussion of the relationship between $\Gamma$ and the blue straggler numbers for the entire cluster, not just the core, can be found in Paper I. 

\begin{figure}
\begin{center}
\includegraphics[width=3.2in]{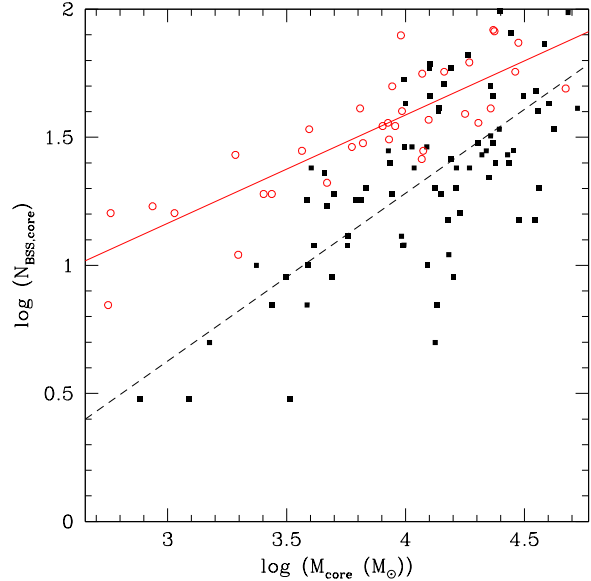}
\end{center}
\caption[]{Number of blue stragglers in the core vs core mass of the
  cluster. The data for model clusters are shown as solid squares, and the
  Milky Way clusters are shown as open circles. The solid line is the line of best fit for the Milky Way clusters, and the dashed line is the best fit to the model clusters}
\label{fig:nmcore}
\end{figure}

The tightest observed correlation between blue straggler populations and cluster properties is between core blue straggler number and core mass, first identified in \citet{Knigge09}. They found a tight but sub-linear correlation, which is reproduced in Figure \ref{fig:nmcore}. Our model clusters match the upper envelope of this correlation very well, but our models show more scatter (a Spearman correlation coefficient of 0.55 compared to 0.83 for the Milky Way data) and an overall slope which is slightly larger than that of the real clusters: 0.6 instead of 0.4 from \citet{Leigh13}. The interpretation of the observational results, specifically the sub-linearity of the correlation, was that the blue stragglers were produced by a process which predominantly depended on binary stars, and did not depend on collisions. We know the detailed formation history of each blue straggler in the models, which are discussed in depth in Paper I. The majority of blue stragglers are produced in binary-mediated collisions. That is to say, a binary star has a strong interaction with a single star or another binary star, and during the course of that interaction, two of the stars physically collide. Both binary stars and collisions are involved, which means we cannot easily separate the two formation mechanisms. 

\begin{figure}
\begin{center}
\includegraphics[width=3.2in]{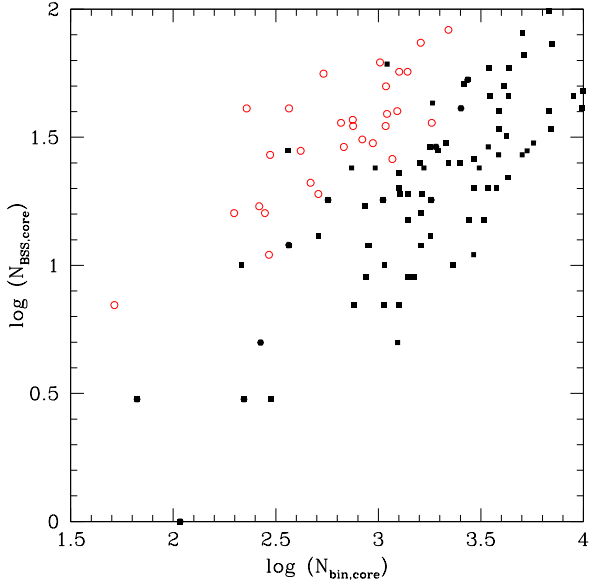}
\end{center}
\caption[]{Number of blue stragglers in the core vs the predicted number of binary stars in the core (from the binary fraction times the core mass). The data for model clusters are shown as solid squares, and the Milky Way clusters are
  shown as open circles. Clusters which were identified as being in the `binary-burning' phase are shown as solid hexagons.}
\label{fig:nnbin}
\end{figure}

In \citet{Leigh13}, the number of blue stragglers was compared to the number of binary stars in the core. The number of binary stars was approximated by multiplying the binary fraction by the core mass, and we have done the same thing here. The binary fractions for the models were determined using a technique that mimics the selection criteria of \citet{Milone12}: we counted the number of main sequence stars between the turnoff and 4 magnitudes fainter, and determined from the models how many of those were binaries with a mass ratio larger than $q = 0.5$. We then doubled that number to get the predicted number of binaries of all mass ratios, as done by \citet{Milone12}. The agreement between the observations and the models is quite good, as seen in figure \ref{fig:nnbin}. A linear fit to both datasets gives the same slope to within 0.003. The models predict slightly fewer blue stragglers, on average, for a given number of binary stars in the core. Since we know that most of the blue stragglers are produced in binary-mediated collisions, it appears that the driving factor in determining whether interactions will produce blue stragglers is dominated by the properties of the binaries, not the number of collisions. Only those interactions with the right combinations of binary masses and orbital parameters will produce blue stragglers. In figure \ref{fig:nnbin} we have identified clusters in the `binary-burning' phase (related to the observational definition of core-collapsed clusters, see \citet{Chatterjee13a} for details) with solid hexagons. The core radii of these clusters may be overestimated since we use the same definition for all clusters, but it is interesting that in or near the core, these clusters follow the same trends as normal clusters. There does not seem to be any substantial modification of either the binary or blue straggler populations around the time of core collapse.

In globular clusters, we have very little observational information about the fraction of blue stragglers which are in binary systems. In open clusters, however, the data are more complete. \citet{Mathieu09} finds that 76$\pm$19\% of blue stragglers are in spectroscopic binaries with periods less than 10$^4$ days in the old open cluster NGC 188, and in the slightly younger cluster M67, 61 $\pm$ 22\% are also binaries with similar periods \citep{Latham96}. This is significantly above the binary fractions for main sequence stars in those clusters (25-30\%). In our models, the blue straggler populations all have binary fractions above 50\%, and can reach as high as 100\%. Therefore, we predict that blue stragglers in clusters should act like their open cluster counterparts and be dominated by binary systems.

\section{Summary and Discussion}

In this paper, we made use of sophisticated Monte Carlo models of globular clusters, which include both dynamical effects and stellar and binary evolution. These models had a realistic range of initial mass, virial radius, binary fraction, and initial concentration, and were evolved to 12 Gyr. We created synthetic color-magnitude diagrams of the evolved clusters, and then determined cluster and blue straggler properties from these color-magnitude diagrams using the techniques that observers use to study the blue straggler populations. 

We find that the algorithm for treating stellar collisions and mergers in the stellar/binary evolution software BSE over-produces the number of bright blue stragglers compared to observations. We confirm that this is a result of the mixing prescription in BSE, and reaffirm that the correction suggested by \citet{Glebbeek08} is more appropriate for correctly modeling blue stragglers. We find that the numbers of blue stragglers selected using observational techniques correlates well with the number selected using a simple mass cutoff. Therefore, we are comfortable using either selection method to study blue straggler populations.

We investigate the correlations between ``observationally" selected blue stragglers and various cluster properties (total mass, core mass, collision rate, binary fraction) and find that the model blue stragglers are consistent with the observed populations. 
Specifically, we find a dependence of blue straggler number on cluster mass, a tighter correlation with core mass, a weak dependence on $\Gamma$, and a stronger dependence on the number of binary stars. The interpretation of the observational results has been that blue stragglers are not collision products but are formed through binary evolution. In the models, the blue stragglers are in fact created in binary-mediated collisions. We need to reconcile this contradiction.

First, more recent observations, combined with careful selection criteria, of blue stragglers show that the number of these objects does show a correlation with collisional parameters, so the idea that blue stragglers cannot be collision products is no longer so clearly ruled out by the observations. However, the dependence of blue straggler number on $\Gamma$ is not as strong as one might expect given the model results which show that the bulk of blue stragglers are formed in binary-mediated collisions. Our understanding of the relationship between the collisional parameter $\Gamma$ and binary interactions is guided by the work of \citet{Leonard89}. He starts with a derivation based on single-single collisions, and calculates the gravitationally focussed cross section for single stars. He then replaces the radius of the star with the semi-major axis of the binary in this calculation to determine the likelihood of a strong interaction between a binary system and another object. These calculations do predict the collision rates of various objects in a cluster, under the assumption that there is a typical stellar mass, binary mass, stellar radius, and binary semi-major axis. However, they do not take into account the possible {\em outcomes} of such interactions. For example, binary-single encounters can produce 11 possible outcomes: preservation, ionization, exchanges, and mergers with different combinations of stars involved \citep{McMillan96} and binary-binary encounters are even more complicated. Only some of those interactions will produce a blue straggler, and the probability of those interactions occurring depend on the properties of the binaries and single stars involved. Even if we have a triple merger, but with three 0.2 \msun~ stars, it will not produce a blue straggler since the mass will still be less than the turnoff mass. Therefore, we conclude that the simple approximations of collision rates in clusters, particularly for binary stars, are not suitable for determining the number of a particular subset of the interactions that occur. To put it another way, the probability that a blue straggler will be formed during a binary-single interaction is more strongly dependent on the binary properties than on the simple collision rate. More accurate analytic calculations for these predictions should, at the very least, include the range of binary properties in the collision rate calculation, and should also include factors which take into account the likelihoods of the various appropriate formation mechanisms (mergers compared to ionizations, for example). 

There are other populations in clusters which are expected to be formed through collisions. In particular, low-mass X ray binaries and cataclysmic binaries are thought to be created when a compact object (a neutron star or a white dwarf) acquires a new binary companion through an exchange interaction or tidal capture, or collides with a giant star. Both these populations {\em do} show a correlation with collision rate in globular clusters \citep{Pooley03, Pooley06, Bahramian13}. We predict that the difference between these populations and the blue straggler populations is that there is a smaller range of binary properties which can produce these populations, so the assumption that the production mechanism is a simple factor of the average encounter rate is more appropriate than for blue stragglers. 

We note that we do not expect to find an exact match between the model blue straggler populations and the sample of blue stragglers in Milky Way clusters. While our model clusters do have global properties which match those of real clusters, we should be careful about making a detailed comparison between the two populations. Our model clusters were chosen to be representative, but we have not attempted to select a population of model clusters with the correct initial conditions of the Milky Way cluster population. For example, we do not draw our model clusters from the present-day cluster mass function for the Milky Way. Similarly, all our clusters were studied at an age of exactly 12 Gyr. The agreement between the blue straggler populations in the models and the Milky Way clusters shown in this paper shows that the blue straggler populations are not very sensitive to these details about the cluster populations. However, if we start looking at more than just overall trends and wish to model the blue straggler population in Milky Way clusters in more detail, we must include a proper population synthesis analysis of the Milky Way clusters.

\section{Acknowledgments}
A.S. is supported by the Natural Sciences and Engineering Research Council of Canada. E. G. is supported by NWO under grant 639.041.129. S.C. acknowledges support from the Department of Astronomy, University of Florida, and also support provided by NASA through a grant from the Space Telescope Science Institute, which is operated by the Association of Universities for Research in Astronomy, Incorporated, under NASA contract NAS5-26555. The grant identifying number is HST-AR-12829.01-A. F. A. R acknowledges support from NASA ATP Grant NNX09AO36G at Northwestern University. All authors would like to thank the Kavli Institute for Theoretical Physics for their hospitality, where this project began. This research was supported in part by the National Science Foundation under Grant No. NSF PHY11-25915.

\end{document}